\documentclass[preprint,prd,aps,showpacs,showkeys,nofootinbib]{revtex4}
\usepackage{graphicx}
\begin{document}
\title{Higgs boson decay $h^0\rightarrow m_VZ$ in the BLMSSM}
\author {Shu-Min Zhao$^1$\footnote{zhaosm@hbu.edu.cn}, Tai-Fu Feng$^{1}$\footnote{fengtf@hbu.edu.cn},
Jian-Bin Chen$^{2}$\footnote{chenjianbin@tyut.edu.cn}, Jing-Jing Feng,
Guo-Zhu Ning$^{1}$\footnote{ninggz@hbu.edu.cn}, Hai-Bin Zhang$^{1}$\footnote{hbzhang@hbu.edu.cn}}
\affiliation{$^1$ Department of Physics, Hebei University, Baoding 071002, China,\\
$^2$ College of Physics and Optoelectronic Engineering,
Taiyuan University of Technology, Taiyuan 030024, China}
\date{\today}

\begin{abstract}
In the framework of BLMSSM, the Higgs decays $h^0\rightarrow Z\gamma$ and $h^0\rightarrow m_VZ$ are
studied where $m_V$ represents a vector meson($\rho, \omega, \phi, J/\psi, \Upsilon$ etc.).
Corrections to the CP-even and CP-odd $h^0\gamma Z$ couplings occur via loop diagrams where new particles are involved.
Actually both of them obviously influence the decay rates of $h^0\rightarrow Z\gamma$ and $h^0\rightarrow m_VZ$.
Concretely, our obtained numerical result shows that the decay width of  $h^0\rightarrow Z\gamma$ can be
1.3 times larger than the  prediction  of the Standard Model(SM).
For the light mesons ($\rho,\omega$), the corrections to $h^0\rightarrow m_VZ$ are within $15\%\sim 20\%$
still consistent with the SM results.
The results of this work would encourage a detection on $h^0\rightarrow Z\gamma$ at LHC for exploring new physics beyond
SM.
\end{abstract}


\maketitle
\section{introduction}
The Higgs boson $h^0$ with mass around 125 GeV was discovered by the CMS and ATLAS collaborations simultaneously
in 2012\cite{higgs125}. As a new elementary particle, $h^0$ is
consistent with the neutral Higgs boson predicted by the Standard Model(SM) to a large extent.
It indeed is a great success because the discovery ends an old epoch and opens a new one.
However, many questions have been raised which challenge the SM framework. If fermions obtain masses uniquely
from their Yukawa couplings with the Higgs field,
the ratio of $m_t/m_e$ is about $3.5\times10^5$;
neutrinos have tiny masses $m_\nu$ around eV order \cite{neutrinomass}.
It is indeed surprising to note the ratio of $m_t/m_\nu$ to be at the order of $2.0\times 10^{11}$\cite{2016pdg}.
If the neutrino masses were simply from the Yukawa couplings with Higgs, one may ask why the gap
is so large? Nowadays it is explained with the so-called see-saw mechanism. The hidden physics scenario is
that the hierarchy problem should be solved in the models beyond SM.

There are many models beyond SM, and almost any of such new models includes more than one
Higgs bosons (charged and neutral)\cite{twoHiggs}. Furthermore, the patterns for the Higgs couplings with
the fermions are more complicated than in SM. Especially, the Higgs couplings which induce flavor changing and CP violating
exist in many of the new models\cite{MSSM,SUSY}. Emergence of those
new particles along with the new interaction causes corrections to
the standard Higgs couplings and may produce non-standard Higgs effective couplings. At tree level,
there is no $h^0\gamma Z$ coupling, but it can be produced by loop diagrams\cite{vertex}. This coupling is very important to probe
the new physics.

In the new models, there are additional charged scalars, vector bosons and fermions coupling with the Higgs boson.
They contribute to the $h^0Z\gamma$
coupling through loop diagrams.
With respect to the SM prediction, the modification of the $h^0Z\gamma$ coupling is expected. To determine whether
the discovered Higgs boson with mass around 125 GeV is indeed the particle in the SM,
it is effective to study $h^0 \rightarrow \gamma\gamma$ and $h^0 \rightarrow Z\gamma$. In the SM prediction, the branching ration of $h^0\rightarrow Z\gamma$  is comparable
with the branching ratio of $h^0\rightarrow \gamma\gamma$ and they are respectively $B(h^0\rightarrow Z\gamma)=(1.54\pm0.09)\times10^{-3}$ \cite{htozgexp}
and $B(h^0\rightarrow \gamma\gamma)=(2.27\pm0.05)\times10^{-3}$.
Taking Higgs boson mass as 125.09 GeV, the ATLAS collaboration give out that
the upper limit on the production cross section times the branching ratio for $pp \rightarrow h^0 \rightarrow Z\gamma$
is 6.6(5.2) times the SM prediction at the 95\% confidence level\cite{htozgexp}.

The authors study the process $h^0\rightarrow m_V\gamma$ in great detail with $m_V$ representing a meson\cite{htomgamma}.
In the work \cite{HQQ1}, the authors use an effective field theory,  where the dimension-six operators correct the Higgs
 couplings to fermions. The dimension-six operators are suppressed by the new physics scale $\Lambda$,
and they can give corrections to the scalar couplings of the Higgs. Beyond SM, there are pseudoscalar couplings of the Higgs,
that are completely the contributions from the dimension-six operators.
 The new physics can give contributions to the dimension-six operators,
and affect the Higgs coupling to quarks\cite{HQQ2}.
With the phenomenological Lagrangian,
the exclusive weak radiative Higgs decays $h^0 \rightarrow m_V V, (V=Z,W)$ are studied as probes for non-standard couplings\cite{htomz}. According to decay topologies,
their contributions are divided into two types: the direct contributions and the indirect contributions.
For the direct contributions, the quarks forming the meson couple to the Higgs boson directly.
On the other hand, the meson is converted by an
off-shell vector boson through the local matrix element\cite{indirect} in the indirect contributions. The direct and indirect
contributions interfere strongly in the decay $h^0\rightarrow m_V\gamma$\cite{htomgamma}. The indirect contributions of
the decay $h^0\rightarrow m_V Z$ are produced from the effective $h^0\gamma Z$ vertex and they are
more important than the direct contributions, especially when $m_V$ is a light vector meson\cite{htomz}.
 QCD factorization \cite{other1,other2} is used for the exclusive weak radiative Higgs decay $h^0\rightarrow m_VZ$.

In the decay $h^0\rightarrow m_V\gamma$, $m_V$ is just a transversely polarized vector meson because of the photon being
transversely polarized.  Since the final state $Z$ boson can be in both longitudinal and transverse polarization states,
the produced mesons can be pseudoscalars and vectors. The effective vertex $h^0\gamma Z$ is important and it obtains
contributions from the new physics through loop diagrams. In this work, we study the Higgs boson
decays $h^0\rightarrow m_VZ$ and $h^0\rightarrow Z\gamma$ in the framework of the BLMSSM
which was first proposed by the authors\cite{BLfirst}. The BLMSSM is the minimal supersymmetric
extension of the SM with local gauged baryon number and lepton number. It can explain the matter-antimatter asymmetry in the Universe.
In this model, Higgs boson mass, the decays $h^0\rightarrow \gamma\gamma$ and
$h^0\rightarrow VV^*, V=(W,Z)$ are researched\cite{BLfirst,TFBL}.
In our previous works,  the lepton flavor violation
processes, lepton EDM, quark EDM are also studied in the BLMSSM\cite{zhaoBL}.

After this introduction, we briefly present the main
ingredients of the BLMSSM in Sec. II.
The Higgs boson decays $h^0\rightarrow Z\gamma$ and $h^0\rightarrow m_VZ$ are studied in Sec. III.  In Sec. IV, we show the input parameters and the numerical results.
The discussion and conclusion are given out  in the last section. Some formulae are shown in the appendix.

\section{The BLMSSM}
\indent\indent
The local gauge group of the BLMSSM is $SU(3)_{C}\otimes SU(2)_{L}\otimes U(1)_{Y}\otimes U(1)_{B}\otimes U(1)_{L}$,
 and the local gauge symmetries are broken through Higgs mechanism.
To cancel the $B$ and $L$ anomalies, the exotic quarks $(\hat{Q}_{4},
\hat{U}_{4}^c, \hat{D}_{4}^c,\hat{Q}_{5}^c,\hat{U}_{5}, \hat{D}_{5})$ and exotic leptons
$(\hat{L}_{4},\hat{E}_{4}^c,\hat{N}_{4}^c,\hat{L}_{5}^c,\hat{E}_{5}, \hat{N}_{5})$ are added.
The Higgs superfields $\hat{\Phi}_{L},\;\hat{\varphi}_{L},\;\hat{\Phi}_{B}$ and $\hat{\varphi}_{B}$ are introduced to provide masses to the exotic leptons and exotic quarks.
 The authors use superfields $\hat{X}$ and $\hat{X}^\prime$ to make the heavy exotic quarks unstable.
 We show these new superfields in Table I.

 \begin{table}
\caption{ The new superfields in the BLMSSM.}
\begin{tabular}{|c|c|c|c|c|c|}
\hline
Superfields & $SU(3)_C$ & $SU(2)_L$ & $U(1)_Y$ & $U(1)_B$ & $U(1)_L$\\
\hline
$\hat{Q}_4$ & 3 & 2 & 1/6 & $B_4$ & 0 \\
$\hat{U}^c_4$ & $\bar{3}$ & 1 & -2/3 & -$B_4$ & 0 \\
$\hat{D}^c_4$ & $\bar{3}$ & 1 & 1/3 & -$B_4$ & 0 \\
$\hat{Q}_5^c$ & $\bar{3}$ & 2 & -1/6 & -$(1+B_4)$ & 0 \\
$\hat{U}_5$ & $3$ & 1 & 2/3 &  $1 + B_4$ & 0 \\
$\hat{D}_5$ & $3$ & 1 & -1/3 & $1 + B_4$ & 0 \\
$\hat{L}_4$ & 1 & 2 & -1/2 & 0 & $L_4$ \\
$\hat{E}^c_4$ & 1 & 1 & 1 & 0 & -$L_4$ \\
$\hat{N}^c_4$ & 1 & 1 & 0 & 0 & -$L_4$ \\
$\hat{L}_5^c$ & 1 & 2 & 1/2 & 0 & -$(3 + L_4)$ \\
$\hat{E}_5$ & 1 & 1 & -1 & 0 & $3 + L_4$ \\
$\hat{N}_5$ & 1 & 1 & 0 & 0 & $3 + L_4$ \\
$\hat{\Phi}_B$ & 1 & 1 & 0 & 1 & 0 \\
$\hat{\varphi}_B$ & 1 & 1 & 0 & -1 & 0 \\
$\hat{\Phi}_L$ & 1 & 1 & 0 & 0 & -2 \\
$\hat{\varphi}_L$ & 1 & 1 & 0 & 0 & 2 \\
$\hat{X}$ & 1 & 1 & 0 & $2/3 + B_4$ & 0 \\
$\hat{X'}$ & 1 & 1 & 0 & $-(2/3 + B_4)$ & 0 \\
$\hat{N}^c$ & 1 & 1 & 0 & 0 & -1 \\
\hline
\end{tabular}
\label{quarks}
\end{table}
$H_{u}$ and $H_{d}$ are $SU(2)_L$ doublets, whose concrete forms are
\begin{eqnarray}
&&H_{u}=\left(\begin{array}{c}H_{u}^+\\{1\over\sqrt{2}}\Big(\upsilon_{u}+H_{u}^0+iP_{u}^0\Big)\end{array}\right)\;,~~~~
H_{d}=\left(\begin{array}{c}{1\over\sqrt{2}}\Big(\upsilon_{d}+H_{d}^0+iP_{d}^0\Big)\\H_{d}^-\end{array}\right)\;.
\label{su2D}
\end{eqnarray}
The $SU(2)_L$ singlets $\Phi_{B},\varphi_{B},\Phi_{L}$ and $
\varphi_{L}$ are written as
   \begin{eqnarray}
&&\Phi_{B}={1\over\sqrt{2}}\Big(\upsilon_{B}+\Phi_{B}^0+iP_{B}^0\Big)\;,~~~~~~~~~
\varphi_{B}={1\over\sqrt{2}}\Big(\overline{\upsilon}_{B}+\varphi_{B}^0+i\overline{P}_{B}^0\Big)\;,
\nonumber\\
&&\Phi_{L}={1\over\sqrt{2}}\Big(\upsilon_{L}+\Phi_{L}^0+iP_{L}^0\Big)\;,~~~~~~~~~~
\varphi_{L}={1\over\sqrt{2}}\Big(\overline{\upsilon}_{L}+\varphi_{L}^0+i\overline{P}_{L}^0\Big)\;.
\label{VEVs}\label{su2S}
\end{eqnarray}
In Eqs.(\ref{su2D}-\ref{su2S}), $\upsilon_{u},\;\upsilon_{d}, \upsilon_{{B}},\overline{\upsilon}_{{B}}, \upsilon_{L}$ and $\overline{\upsilon}_{L}$
are all nonzero vacuum expectation values VEVs.

The superpotential of the BLMSSM \cite{BLfirst,TFBL} reads as
\begin{eqnarray}
&&{\cal W}_{{BLMSSM}}={\cal W}_{{MSSM}}+{\cal W}_{B}+{\cal W}_{L}+{\cal W}_{X}\;,
\label{superpotential1}
\nonumber\\&&{\cal W}_{B}=\lambda_{Q}\hat{Q}_{4}\hat{Q}_{5}^c\hat{\Phi}_{B}+\lambda_{U}\hat{U}_{4}^c\hat{U}_{5}
\hat{\varphi}_{B}+\lambda_{D}\hat{D}_{4}^c\hat{D}_{5}\hat{\varphi}_{B}+\mu_{B}\hat{\Phi}_{B}\hat{\varphi}_{B}
\nonumber\\
&&\hspace{1.2cm}
+Y_{{u_4}}\hat{Q}_{4}\hat{H}_{u}\hat{U}_{4}^c+Y_{{d_4}}\hat{Q}_{4}\hat{H}_{d}\hat{D}_{4}^c
+Y_{{u_5}}\hat{Q}_{5}^c\hat{H}_{d}\hat{U}_{5}+Y_{{d_5}}\hat{Q}_{5}^c\hat{H}_{u}\hat{D}_{5}\;,
\nonumber\\
&&{\cal W}_{L}=Y_{{e_4}}\hat{L}_{4}\hat{H}_{d}\hat{E}_{4}^c+Y_{{\nu_4}}\hat{L}_{4}\hat{H}_{u}\hat{N}_{4}^c
+Y_{{e_5}}\hat{L}_{5}^c\hat{H}_{u}\hat{E}_{5}+Y_{{\nu_5}}\hat{L}_{5}^c\hat{H}_{d}\hat{N}_{5}
\nonumber\\
&&\hspace{1.2cm}
+Y_{\nu}\hat{L}\hat{H}_{u}\hat{N}^c+\lambda_{{N^c}}\hat{N}^c\hat{N}^c\hat{\varphi}_{L}
+\mu_{L}\hat{\Phi}_{L}\hat{\varphi}_{L}\;,
\nonumber\\
&&{\cal W}_{X}=\lambda_1\hat{Q}\hat{Q}_{5}^c\hat{X}+\lambda_2\hat{U}^c\hat{U}_{5}\hat{X}^\prime
+\lambda_3\hat{D}^c\hat{D}_{5}\hat{X}^\prime+\mu_{X}\hat{X}\hat{X}^\prime.
\label{superpotential-BL}
\end{eqnarray}
Here ${\cal W}_{{MSSM}}$ is the superpotential of the MSSM.
To save space in the text, we do not show the soft breaking terms $\mathcal{L}_{{soft}}$ here,
which can be found in the previous work\cite{TFBL}.

\section{the processes $h^0\rightarrow m_V Z$ and $h^0\rightarrow Z\gamma$}
\begin{figure}[h]
\setlength{\unitlength}{1mm}
\centering
\includegraphics[width=1.4in]{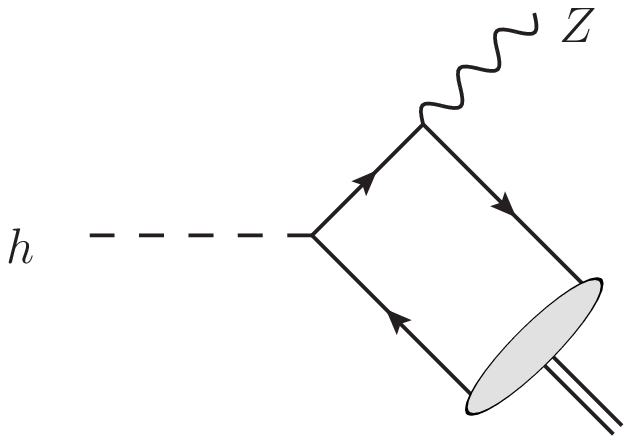}~\includegraphics[width=1.4in]{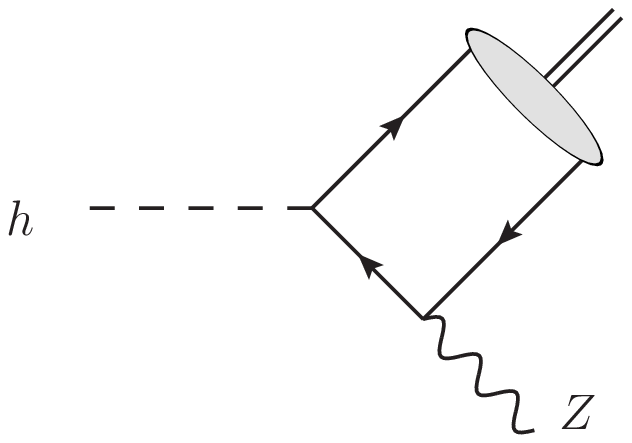}
~\includegraphics[width=1.4in]{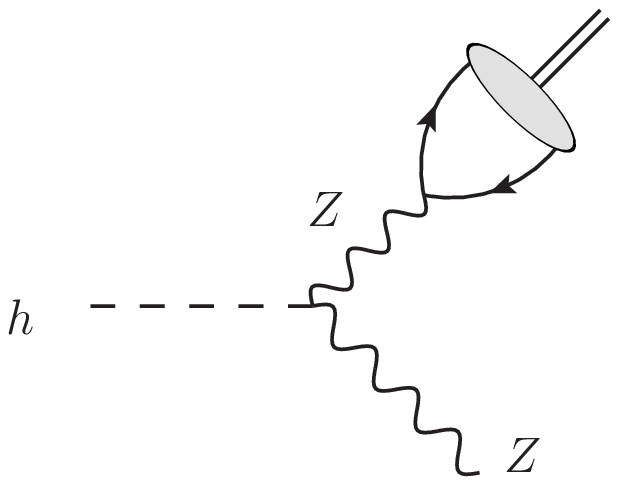}~\includegraphics[width=1.4in]{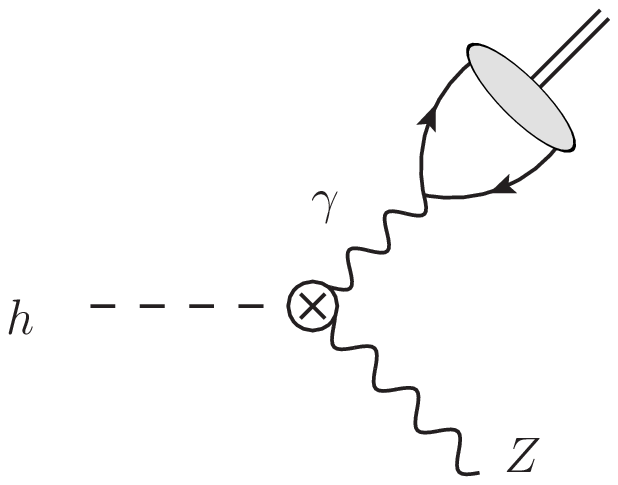}
\caption[]{The diagrams contributing to the decay $h^0\rightarrow m_VZ$.
The crossed circle in the last graph represents the effective vertex
$h^0\rightarrow Z\gamma^*$ from the one loop diagrams.}\label{htoMVZ}
\end{figure}

For the Higgs boson weak hadronic decay $h^0\rightarrow m_V Z$, the direct contributions are
described by the first two  diagrams in FIG.{\ref{htoMVZ}}. The quark and anti-quark forming the final state meson
couple to the Higgs boson directly. As discussed detailedly in the Ref.\cite{htomz}, the contributions from the first two diagrams in FIG.{\ref{htoMVZ}} are not dominant, though they are tree diagrams.
 In FIG.1, the last two diagrams represent the indirect contributions.
In the process $h^0\rightarrow ZZ^*\rightarrow m_V Z$, $Z^*$ is off-shell and changes into the final state meson.
$h^0\rightarrow ZZ^*$ can occur at tree level in the SM. The tree level vertex $h^0\gamma Z$ does not exist, but it can be produced
through loop diagrams. In the BLMSSM, the non-standard $h^0\gamma Z$ vertex should be taken into account.
The effective Lagrangian for $h^0\gamma Z$ is written in the following form
\begin{eqnarray}
\mathcal{L}_{eff}=\frac{\alpha}{4\pi \upsilon}\Big(\frac{2C_{\gamma Z}}{s_Wc_W}hF_{\mu\nu}Z^{\mu\nu}
-\frac{2\tilde{C}_{\gamma Z}}{s_Wc_W}hF_{\mu\nu}\tilde{Z}^{\mu\nu}  \Big),\label{HZgammaE}
\end{eqnarray}
with $s_W=\sin\theta_W, ~c_W=\cos\theta_W$. Here, $\theta_W$ is the weak mixing angle and $\upsilon=\sqrt{\upsilon_u^2+\upsilon_d^2}$.
Using the effective Lagrangian in Eq.(\ref{HZgammaE}), we show the decay width of $h^0\rightarrow Z\gamma$
\begin{eqnarray}
\Gamma(h^0\rightarrow Z\gamma)=\frac{\alpha^2m_{h^0}^3}{32\pi^3\upsilon^2s_W^2c_W^2}(1-\frac{m_Z^2}{m_{h^0}^2})^3(|C_{\gamma Z}|^2+|\tilde{C}_{\gamma Z}|^2).
\end{eqnarray}

The loop diagrams with new physics can produce additional
corrections to both $h^0\rightarrow m_V \gamma$ and $h^0\rightarrow  m_V Z$.
Though the processes $h^0\rightarrow m_V \gamma$ and $h^0\rightarrow  m_V Z$ seem similar, they have essential difference.
For $h^0\rightarrow  m_V \gamma$, the final state photon is on shell and massless, which leads to the loss of longitudinal polarization.
We show the invariant matrix element for $h^0\rightarrow m_V \gamma$ at tree level in NRQCD
\begin{eqnarray}
&&{\cal M}^{\gamma} = \frac{4\sqrt{3} e e_q\phi_0}{m^2_{h^0} - m^2_V}\Big(\frac{m_VG_F}{2\sqrt{2}}\Big)^{\frac{1}{2}}
[c_S\{2(\varepsilon^*_\gamma\cdot p_V)(\varepsilon^*_V\cdot k_\gamma) - (m^2_{h^0} - m^2_V)(\varepsilon^*
_\gamma\cdot\varepsilon^*_V)\}\nonumber\\
&&-2c_P\epsilon_{\mu\nu\rho\lambda}~\varepsilon^{*\mu}_\gamma k_\gamma^\nu p^\rho
_V\varepsilon^{*\lambda}_V]\label{gs1}.
\end{eqnarray}
Here, $k_\gamma(p_V)$ is the four-momentum of the photon$(m_V)$, while $\varepsilon^*_\gamma(\varepsilon^*_V)$ is the
polarization of the photon$(m_V)$.

In the rest frame of the $m_V$, Eq.(\ref{gs1}) can be written in a more familiar form. With the
definition $\varepsilon^{*L}_V\equiv \bar{\varepsilon}^{*}_V\cdot\hat{k}_\gamma$ and $\bar{\varepsilon}^{*T}_V\equiv \bar{\varepsilon}^{*}_V-\varepsilon^{*L}_V\hat{k}_\gamma$,
we obtain the following formula in the transverse basis\cite{HQQ1}
\begin{eqnarray}
&&{\cal M}^\gamma = H^{\gamma}_{\parallel}{\vec\varepsilon}^{*T}_V\cdot\vec\varepsilon^*_\gamma + iH^{\gamma}_\perp
{\hat k}_\gamma\cdot({\vec\varepsilon}^{*T}_V\times\vec\varepsilon^*_\gamma), \label{gs2}\\&&
H^{\gamma}_{\parallel}= 4\sqrt{3} e e_q\phi_0\Big(\frac{m_VG_F}{2\sqrt{2}}\Big)^{\frac{1}{2}} c_S,\nonumber\\
&& H^{\gamma}_\perp= 4\sqrt{3} e e_q\phi_0\Big(\frac{m_VG_F}{2\sqrt{2}}\Big)^{\frac{1}{2}} ic_P.\nonumber
\label{Hperpdef}
\end{eqnarray}
Eq.(\ref{gs2}) does not include longitudinal polarization, because the on-shell photon is massless and has no
longitudinal polarization. The triple product ${\hat k}_\gamma\cdot({\vec\varepsilon}^{*T}_V\times\vec\varepsilon^*_\gamma)$
is the only P-odd observable in $|\mathcal{M}|^2$, and its coefficient is proportional to $c_Sc_P$.
$c_P$ is the new physics pseudoscalar $Hq\bar{q}$ coupling, which is embodied by the nonzero value of the triple product.
Unfortunately, the photon dose not decay and we are unable to determine $\varepsilon^*_\gamma$, that leads to the failure for the
measurement of $c_P$.

To solve this problem, we replace the photon with a vector boson $Z$, whose polarization can be measured through its decay.
Different from the photon, there is tree level coupling of $Z$ and Higgs boson. Therefore, an additional tree-level diagram contributes to
the decay $h^0\rightarrow m_V Z$. The coupling of $Z\bar{q}q$ includes axial-vector term,
but this term in the matrix element of $h^0\rightarrow m_V Z$ vanishes at the leading order in NRQCD.
We take $\hat{k}_Z$ as the direction of $Z$ in the rest frame of $m_V$.
Using the similar analysis as $h^0\rightarrow m_V \gamma$, we obtain the formula
in the rest frame of $m_V$ and the transverse basis
\begin{eqnarray}
&&{\cal M}^Z = H^Z_{0}{\vec\varepsilon}^{*L}_V\cdot\vec\varepsilon^{*L}_Z+H^Z_{\parallel}{\vec\varepsilon}^{*T}_V\cdot\vec\varepsilon^{*T}_Z + iH^Z_\perp
{\hat k}_Z\cdot({\vec\varepsilon}^{*T}_V\times\vec\varepsilon^{*T}_Z).
\end{eqnarray}
$H_0^Z, ~H^Z_{\parallel}$ are in direct proportion to $c_S$ and $H^Z_\perp$ is in direct proportion to $c_P$. The concrete forms of
$H_0^Z, ~H^Z_{\parallel}$ and $c_P^Z$ can be found in Ref.\cite{HQQ1}. Through the decay products of $Z$, people can measure $\varepsilon^{*T}_Z.$
If nonzero value of triple product is measured, one can be convinced of a clear signal of $c_P$.

For the decay $h^0\rightarrow m_V Z$,
the Feynman amplitudes are generally parameterized as
\begin{equation}
   i{\cal A}(h^0\to m_VZ)
   = - \frac{2g m_V}{c_W \upsilon}
    \left[ \varepsilon_V^{\parallel *}\cdot\varepsilon_Z^{\parallel *}\,F_\parallel^{VZ}
    + \varepsilon_V^{\perp *}\cdot\varepsilon_Z^{\perp *}\,F_\perp^{VZ}
    + \frac{\epsilon_{\mu\nu\alpha\beta}\,k_V^\mu k_Z^\nu\varepsilon_V^{*\alpha}\varepsilon_Z^{*\beta}}%
           {\left[ (k_V\cdot k_Z)^2-k_V^2 k_Z^2\right]^{1/2}}\,\widetilde F_\perp^{VZ} \right] ,
\end{equation}
with $k_Z$ representing the momentum of $Z$. $\varepsilon_V^{\parallel \mu}$
is the longitudinal polarization vector of the meson, while $\varepsilon_V^{\perp\mu}$ denotes the transverse polarization
vector\cite{htomz}
\begin{equation}
   \varepsilon_V^{\parallel \mu} = \frac{1}{m_V}\,\frac{k_V\cdot k_Z}{\left[ (k_V\cdot k_Z)^2-k_V^2 k_Z^2\right]^{1/2}}
    \left( k_V^\mu - \frac{k_V^2}{k_V\cdot k_Z}\,k_Z^\mu \right) , \qquad
   \varepsilon_V^{\perp\mu} = \varepsilon_V^\mu - \varepsilon_V^{\parallel \mu}.\label{vectorV}
\end{equation}
In Eq.(\ref{vectorV}), one obtains the polarization vectors of $Z$ using the replacement $m_V\rightarrow m_Z, k_V\leftrightarrow k_Z$.

The decay width of $h^0\to m_VZ$ is expressed in the following form
\begin{eqnarray}
   &&\Gamma(h^0\to m_VZ)
   = \frac{m_{h^0}^3}{4\pi \upsilon^4}\,\lambda^{1/2}(1,r_Z,r_V)\,(1-r_Z-r_V)^2 \nonumber \\&&
   \quad\times \left[
    \big| F_\parallel^{VZ} \big|^2 + \frac{8r_V r_Z}{(1-r_Z-r_V)^2}
    \Big( \big| F_\perp^{VZ} \big|^2 + \big| \widetilde F_\perp^{VZ} \big|^2 \Big) \right] ,
\end{eqnarray}
with $\lambda(x,y,z)=(x-y-z)^2-4yz$, $r_Z=m_Z^2/m_{h^0}^2$ and $r_V=m_V^2/m_{h^0}^2$. For light vector mesons, the mass
ratios $r_V=m_V^2/m_{h^0}^2$ are small, but we keep them in our study for better results.
\begin{eqnarray}
  &&   F_{\parallel\,\rm indirect}^{VZ} = \frac{\kappa_Z}{1-r_V/r_Z} \sum_q f_V^q\,v_q
    + C_{\gamma Z}\,\frac{\alpha(m_V)}{4\pi}\,\frac{4r_Z}{1-r_Z-r_V} \sum_q f_V^q\,Q_q,  \nonumber\\&&
   F_{\perp\,\rm indirect}^{VZ} = \frac{\kappa_Z}{1-r_V/r_Z} \sum_q f_V^q\,v_q
    + C_{\gamma Z}\,\frac{\alpha(m_V)}{4\pi}\,\frac{1-r_Z-r_V}{r_V} \sum_q f_V^q\,Q_q, \nonumber\\&&
   \widetilde F_{\perp\,\rm indirect}^{VZ}
   = \widetilde C_{\gamma Z}\,\frac{\alpha(m_V)}{4\pi}\,
    \frac{\lambda^{1/2}(1,r_Z,r_V)}{r_V} \sum_q f_V^q\,Q_q.\label{FVZindirect}
\end{eqnarray}
Here, the vector and axial-vector couplings of $Z\bar{q}q$ are denoted respectively by
$v_q=\frac{T^q_3}{2}-Q_qs_W^2$ and $a_q=\frac{T^q_3}{2}$. $f_V^q$ is the vector meson decay constant,
whose definition reads as
\begin{eqnarray}
&&\langle V(k,\varepsilon)|\bar{q}\gamma^\mu q |0\rangle=-if^q_Vm_V\varepsilon^{*\mu},~~~~~~q=u,d,s\dots
\end{eqnarray}
To calculate the results, the following relations are used
\begin{eqnarray}
&&Q_Vf_V=\sum_qQ_qf^q_V, ~~~~~~~\sum_qf^q_Vv_q=f_Vv_V.
\end{eqnarray}
The concrete forms of $C_{\gamma Z}$ and $\widetilde C_{\gamma Z}$ in Eq.(\ref{FVZindirect}) are shown here\cite{CGZ}
\begin{eqnarray}
   &&C_{\gamma Z}= C_{\gamma Z}^{SM}+C_{\gamma Z}^{New}, ~~~~~~~~~~~~~~   \widetilde C_{\gamma Z}
   = \widetilde C_{\gamma Z}^{SM}+\widetilde C_{\gamma Z}^{New}.\nonumber\\
   &&C_{\gamma Z}^{SM}= \sum_q \frac{2 N_c Q_q v_q}{3}\,A_f(\tau_q,r_Z)
    + \sum_l \frac{2 Q_l v_l}{3}\,A_f(\tau_l,r_Z)
    - \frac{1}{2}\,A_W^{\gamma Z}(\tau_W,r_Z),
 \end{eqnarray}
where $\tau_i=\frac{4m_i^2}{m_{h^0}^2}$. $C_{\gamma Z}^{SM}$ and $\widetilde C_{\gamma Z}^{SM}$ represent the
SM contributions to $h^0\rightarrow Z\gamma$. $A_f, B_f$ and $A_W^{\gamma Z}$ are all loop functions\cite{htomgamma}. Using the running quark mass and the low-energy
values given in Ref.\cite{PDG2014}, the authors give out the numerical values of $C_{\gamma Z}^{SM}$ and $\widetilde C_{\gamma Z}^{SM}$:
 $C_{\gamma Z}^{SM}\sim-2.395+0.001i,~~\widetilde C_{\gamma Z}^{SM}\sim 0$.

\begin{figure}[h]
\setlength{\unitlength}{1mm}
\centering
\includegraphics[width=6in]{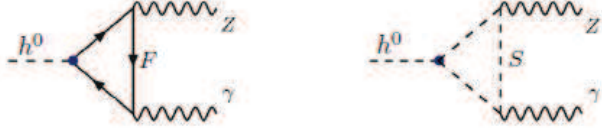},
\vspace{-17cm}
\caption[]{The one loop diagrams with new particles for $h^0\rightarrow Z\gamma$.}\label{OLDhzg}
\end{figure}

 In the BLMSSM, the new physics one loop diagrams for $h^0\rightarrow Z\gamma$ are shown in FIG.\ref{OLDhzg}, with $F$ denoting charged Fermions and $S$ denoting charged scalars.
 The new contributions to $C_{\gamma Z}$
 originate from the exchanged particles: exotic leptons, exotic quarks, charginos, sleptons, squarks, exotic sleptons, exotic squarks and charged Higgs.
 \begin{eqnarray}
&&C^{New}_{\gamma Z}=\frac{ \upsilon s_Wc_W }{e}\int_0^1dx\int_0^{1-x}dy\Big\{\nonumber\\&&
\sum_{F=l',b',t',\chi^{\pm}}\Big[\frac{Q_{F_1}}{R_1^2(m_{F_1},m_{F_2})}\Big(m_{F_1}(2x-1)y(A^{\bar{F}_2F_1h^0}_wB^{\bar{F}_1F_2h^0}_w-A^{\bar{F}_2F_1h^0}B^{\bar{F}_1F_2h^0})
\nonumber\\&&-m_{F_2}(A_w^{\bar{F}_2F_1h^0}B_w^{\bar{F}_1F_2h^0}+A^{\bar{F}_2F_1h^0}B^{\bar{F}_1F_2h^0})(2xy+y-1)\Big)
\nonumber\\&&+\frac{Q_{F_1}}{R_2^2(m_{F_1},m_{F_2})}\Big(
m_{F_1}(A^{\bar{F}_2F_1h^0}_wB^{\bar{F}_1F_2h^0}_w-A^{\bar{F}_2F_1h^0}B^{\bar{F}_1F_2h^0})(2xy+y-1)\nonumber\\&&
+m_{F_2}(A_w^{\bar{F}_2F_1h^0}B^{\bar{F}_1F_2h^0}_w
+A^{\bar{F}_2F_1h^0}B^{\bar{F}_1F_2h^0})(1-2x)y\Big)\Big]\nonumber\\&&
+\sum_{S=\tilde{L},\tilde{D},\tilde{U},\tilde{L}',\tilde{D}',\tilde{U}',H^{\pm}}Q_{S_1}A^{S^*_2S_1h^0}B^{S^*_1S_2h^0}
\Big(\frac{xy}{R_1^2(m_{S_1},m_{S_2})}+\frac{xy}{R_2^2(m_{S_1},m_{S_2})}\Big)\Big\}.
\end{eqnarray}

The functions $R_1^2(m_1,m_2)$ and $R_2^2(m_1,m_2)$ read as
\begin{eqnarray}
&&R_1^2(m_1,m_2)=m_1^2\Big(y+\frac{m_2^2}{m_1^2}(1-y)
-\frac{2p_1.p_2}{m_1^2}xy+
\frac{M^2}{m_1^2}x(x-1)+\frac{m_Z^2}{m_1^2}y(y-1)\Big),\nonumber\\&&
R_2^2(m_1,m_2)=m_1^2\Big(1-y+\frac{m_2^2}{m_1^2}y
-\frac{2p_1.p_2}{m_1^2}xy+
\frac{M^2}{m_1^2}x(x-1)+\frac{m_Z^2}{m_1^2}y(y-1)\Big).
\end{eqnarray}
$A^{\bar{F}_2F_1h^0}$ and $A_w^{\bar{F}_2F_1h^0}$ are coupling constants for the vertex $\bar{F}_2F_1h^0$
which is written in the general form $\bar{F}_2i(A^{\bar{F}_2F_1h^0}+A_w^{\bar{F}_2F_1h^0}\gamma_5)F_1h^0$.
In the same way, we use $i\bar{F}_1(B^{\bar{F}_1F_2Z}\gamma_\mu+B_w^{\bar{F}_1F_2Z}\gamma_{\mu}\gamma_5)F_2Z^{\mu}$ for the vertex $\bar{F}_1F_2Z$.

The SM value of $\tilde{C}_{\gamma Z}$ is zero, therefore only new physics contribute to  $\tilde{C}_{\gamma Z}$.
In the BLMSSM, these new particles giving corrections to $\tilde{C}_{\gamma Z}$ are exotic leptons, exotic quarks, charginos
in the one loop diagrams. The scalar loop represented by the right diagram in FIG.2 does not contribute to the CP-odd coupling.
\begin{eqnarray}
&&\tilde{C}^{New}_{\gamma Z}=\frac{i \upsilon s_Wc_WQ_{S_1}}{e }\int_0^1dx\int_0^{1-x}dy
\nonumber\\&&\times\sum_{F=l',b',t',\chi^{\pm}}
\Big[\frac{1}{R_1^2(m_{F_1},m_{F_2})}
\Big(m_{F_1}(A^{\bar{F}_2F_1h^0}_wB^{\bar{F}_1F_2h^0}-A^{\bar{F}_2F_1h^0}B^{\bar{F}_1F_2h^0}_w)y
\nonumber\\&&+m_{F_2}(A^{\bar{F}_2F_1h^0}B^{\bar{F}_1F_2h^0}_w+A^{\bar{F}_2F_1h^0}_wB^{\bar{F}_1F_2h^0})(1-y)\Big)\nonumber\\&&
+\frac{1}{R_2^2(m_{F_1},m_{F_2})}
\Big(m_{F_1}(A^{\bar{F}_2F_1h^0}_wB^{\bar{F}_1F_2h^0}-A^{\bar{F}_2F_1h^0}B^{\bar{F}_1F_2h^0}_w)(1-y)\nonumber\\&&
+m_{F_2}(A^{\bar{F}_2F_1h^0}B^{\bar{F}_1F_2h^0}_w+A^{\bar{F}_2F_1h^0}_wB^{\bar{F}_1F_2h^0})y\Big)\Big].
\end{eqnarray}
In the following, we show the concrete forms of the needed couplings
$A^{\bar{F}_2F_1h^0}$, $A_w^{\bar{F}_2F_1h^0},~
B^{\bar{F}_1F_2h^0},~B_w^{\bar{F}_1F_2h^0}$.
 \begin{eqnarray}
  &&\mathcal{L}_{h^{0}L^\prime L^\prime}=
 \sum_{i,j=1}^2
  \overline{L}^{\prime}_{i+3}\Big[\frac{1}{2}\Big(\frac{Y_{e_4}}{\sqrt{2}}(W_{L}^\dag)^{i2}U_L^{1j}\sin\alpha+\frac{Y_{e_5}}
  {\sqrt{2}}(W_{L}^\dag)^{i1}U_L^{2j}\cos\alpha\nonumber\\&&\hspace{1.6cm}
  +\frac{Y_{e_4}^*}{\sqrt{2}}W_{L}^{2j}(U_L^\dag)^{i1}
  \sin\alpha+\frac{Y_{e_5}^*}{\sqrt{2}}W_{L}^{1j}(U_L^\dag)^{i2}\cos\alpha\Big)
  \nonumber\\&&\hspace{1.6cm}
  +\frac{1}{2}\Big(\frac{Y_{e_4}^*}{\sqrt{2}}W_{L}^{2j}(U_L^\dag)^{i1}
  \sin\alpha+\frac{Y_{e_5}^*}{\sqrt{2}}W_{L}^{1j}(U_L^\dag)^{i2}\cos\alpha\nonumber\\&&\hspace{1.6cm}
  -\frac{Y_{e_4}}{\sqrt{2}}(W_{L}^\dag)^{i2}U_L^{1j}\sin\alpha
  -\frac{Y_{e_5}}  {\sqrt{2}}(W_{L}^\dag)^{i1}U_L^{2j}\cos\alpha\Big)\gamma_5\Big]L^\prime_{j+3}h^0.
  \end{eqnarray}
  The couplings for the Higgs boson $h^0$ and exotic quarks are deduced in Ref.\cite{TFBL},
\begin{eqnarray}
&&{\cal L}_{h^0q' q'}=\sum\limits_{\alpha=1}^8\sum\limits_{i,j=1}^2\Big\{
h^0\overline{t}_{i+3}\Big[\Big((\mathcal{N}^L_{h^0})_{ij}+(\mathcal{N}^R_{h^0})_{ij}
\Big)+\Big((\mathcal{N}^R_{h^0})_{ij}-(\mathcal{N}^L_{h^0})_{ij}
\Big)\gamma_5\Big]t_{j+3}
\nonumber\\&&\hspace{1.8cm}+h^0\overline{b}_{i+3}
\Big[\Big((\mathcal{K}^L_{h^0})_{ij}+(\mathcal{K}^R_{h^0})_{ij}
\Big)+\Big((\mathcal{K}^R_{h^0})_{ij}-(\mathcal{K}^L_{h^0})_{ij}
\Big)\gamma_5\Big]b_{j+3}\Big\}.
\label{H0qq}
\end{eqnarray}
  The coupling constants $(\mathcal{N}^L_{h^0})_{ij}, (\mathcal{N}^R_{h^0})_{ij}, (\mathcal{K}^L_{h^0})_{ij}$
 and $(\mathcal{K}^R_{h^0})_{ij}$ are
\begin{eqnarray}
&&(\mathcal{N}^L_{h^0})_{ij}=\frac{1}{2\sqrt{2}}\Big[Y_{u_4}(W_{t}^\dagger)_{i2}(U_{t})_{1j}\cos\alpha
+Y_{u_5}(W_{t}^\dagger)_{i1}(U_{t})_{2j}\sin\alpha\Big],\nonumber\\&&
(\mathcal{N}^R_{h^0})_{ij}=\frac{1}{2\sqrt{2}}\Big[Y^*_{u_4}(U_{t}^\dagger)_{i1}(W_{t})_{2j}\cos\alpha
+Y^*_{u_5}(U_{t}^\dagger)_{i2}(W_{t})_{1j}\sin\alpha\Big].
\nonumber\\
&&(\mathcal{K}^L_{h^0})_{ij}=\frac{1}{2\sqrt{2}}\Big[Y_{d_4}(W_{b}^\dagger)_{i2}(U_{b})_{1j}\sin\alpha
-Y_{d_5}(W_{b}^\dagger)_{i1}(U_{b})_{2j}\cos\alpha\Big],\nonumber\\&&
(\mathcal{K}^R_{h^0})_{ij}=\frac{1}{2\sqrt{2}}[Y^*_{d_4}(U_{b}^\dagger)_{i1}(W_{b})_{2j}\sin\alpha
-Y^*_{d_5}(U_{b}^\dagger)_{i2}(W_{b})_{1j}\cos\alpha\Big].
\end{eqnarray}

One neutral vector boson($\gamma , Z$) couples to exotic leptons\cite{SM14JHEP}
 \begin{eqnarray}
  &&\mathcal{L}_{VL^\prime L^\prime}
=\sum_{i,j=1}^2
\Big\{ eZ_{\mu}\overline{L}_{i+3}
  \Big[\Big(-\frac{s_W}{c_W}\delta_{ij}
  +\frac{(U_{L}^\dag)^{i1}U_{L}^{1j}+(W_{L}^\dag)^{i1}W_{L}^{1j}}{4s_Wc_W}\Big)\gamma^{\mu}\nonumber\\&&+
  \Big(\frac{(W_{L}^\dag)^{i1}W_{L}^{1j}-(U_{L}^\dag)^{i1}U_{L}^{1j}}{4s_Wc_W}\Big)\gamma^{\mu}\gamma_5\Big]L^\prime_{j+3}
\Big\}+\sum_{i=1}^2eF_{\mu}\overline{L}^\prime_{i+3}\gamma^{\mu}L^\prime_{i+3}+h.c.
  \end{eqnarray}

We show one neutral vector boson($\gamma , Z$) coupling to exotic sleptons
\begin{eqnarray}
&&\mathcal{L}_{V\tilde{L}^\prime \tilde{L}^\prime}=eF_{\mu}\sum_{i,j=1}^2\tilde{E}^{\prime i*}_4i\tilde{\partial}^{\mu}\tilde{E}^{\prime j}_4\delta^{ij}+
eZ_{\mu}\sum_{i,j=1}^2[-\frac{s_W}{c_W}\delta^{ij}+\frac{(Z_{\tilde{e}_4}^\dag)^{i1}Z_{\tilde{e}_4}^{1j}}{2s_Wc_W}]\tilde{E}^{\prime i*}_4i\tilde{\partial}^{\mu}\tilde{E}^{\prime j}_4\nonumber\\&&
+eF_{\mu}\sum_{i,j=1}^2\tilde{E}^{\prime i*}_5i\tilde{\partial}^{\mu}\tilde{E}^{\prime j}_5\delta^{ij}+
eZ_{\mu}\sum_{i,j=1}^2[-\frac{s_W}{c_W}\delta^{ij}+\frac{(Z_{\tilde{e}_5}^\dag)^{i2}Z_{\tilde{e}_5}^{2j}}{2s_Wc_W}]\tilde{E}^{\prime i*}_5i\tilde{\partial}^{\mu}\tilde{E}^{\prime j}_5
+h.c.\label{VSSLPLP}
\end{eqnarray}

The Lagrangian for one neutral vector boson($\gamma, Z$) and exotic squarks couplings are \cite{SM14JHEP}
\begin{eqnarray}
&&\mathcal{L}_{V\tilde{\mathcal{Q}}\tilde{\mathcal{Q}}}=-\frac{2}{3}e\sum_{j,\beta=1}^4\delta_{j\beta}F_{\mu}
\tilde{\mathcal{U}}_{j}^*i\tilde{\partial}^{\mu}
\tilde{\mathcal{U}}_{\beta}
+\frac{e}{3}\sum_{j,\beta=1}^4\delta_{j\beta}F_{\mu}
\tilde{\mathcal{D}}_{j}^*i\tilde{\partial}^{\mu}
\tilde{\mathcal{D}}_{\beta}\nonumber\\&&+\frac{e}{6s_Wc_W}\sum_{j,\beta=1}^4
\Big(4s^2_W\delta_{j\beta}-3(U^{\dag}_{j1}U_{1\beta}+U^{\dag}_{j3}U_{3\beta})\Big)Z_{\mu}
\tilde{\mathcal{U}}_{j}^*i\tilde{\partial}^{\mu}
\tilde{\mathcal{U}}_{\beta}\nonumber\\&&+\frac{e}{6s_Wc_W}\sum_{j,\beta=1}^4
\Big(-2s^2_W\delta_{j\beta}+3(D^{\dag}_{j1}D_{1\beta}+D^{\dag}_{j3}D_{3\beta})\Big)Z_{\mu}
\tilde{\mathcal{D}}_{j}^*i\tilde{\partial}^{\mu}
\tilde{\mathcal{D}}_{\beta}
+h.c.\label{VSSQpQP}
\end{eqnarray}
The neutral vector bosons couple to the exotic quarks
\begin{eqnarray}
&&\mathcal{L}_{ V\mathcal{Q} \mathcal{Q}}
=
-\frac{2e}{3}F_{\mu}\sum_{i=1}^2
\bar{t}_{i+3}\gamma^{\mu}t_{i+3}+\frac{e}{3}F_{\mu}\sum_{i=1}^2\bar{b}_{i+3}\gamma^{\mu}b_{i+3}\nonumber
\\&&+\frac{e}{12s_{W}c_{W}}Z_{\mu}\sum_{j,k=1}^2\bar{t}_{j+3}
\Big[\Big(3(W_t^{\dag})_{j2}(W_t)_{2k}-3(U_t^{\dag})_{j2}(U_t)_{2k}\Big)\gamma^{\mu}\gamma_5\nonumber\\&&
+\Big(2(1-4c_{W}^2)\delta_{jk}
+3(U_t^{\dag})_{j2}(U_t)_{2k}+3(W_t^{\dag})_{j2}(W_t)_{2k}\Big)\gamma^{\mu}
\Big]t_{k+3}\nonumber\\&&
+\frac{e}{12s_{W}c_{W}}Z_{\mu}\sum_{j,k=1}^2\bar{b}_{j+3}\Big[\Big(3(U_b^{\dag})_{j2}(U_b)_{2k}
-3(W_b^{\dag})_{j2}(W_b)_{2k}\Big)\gamma^{\mu}\gamma_5\nonumber\\&&+
\Big(2(1+2c_{W}^2)\delta_{jk}
-3(U_b^{\dag})_{j2}(U_b)_{2k}-3(W_b^{\dag})_{j2}(W_b)_{2k}\Big)\gamma^{\mu}
\Big]b_{k+3}+h.c.\label{VQQ}
\end{eqnarray}
To save space in the text, the couplings for $h^0$-exotic
slepton-exotic slepton and $h^0$-exotic squark-exotic squark are collected in the Appendix that includes the couplings $A^{S^*_2S_1h^0}$ and $B^{S^*_1S_2h^0}$.

As discussed by the authors, the QCD corrections to the
process $h^0\rightarrow Z\gamma$ are around $0.1\%$\cite{QCDCR}. That is to say, the QCD corrections are very small and can be
neglected safely. Up to now, LHC have not observed the decay $h^0\rightarrow Z\gamma$. Announced by CMS and ATLAS, the upper bound on this decay is about
six times the SM results, at 95\% confidence level\cite{UPbound}. The constraint for the parameters is obtained
from the bound and shown as
\begin{eqnarray}
\sqrt{|C_{\gamma Z}|^2+|\tilde{C}_{\gamma Z}|^2}<4.76.
\end{eqnarray}
For the light vector mesons, the contributions from the photon-pole diagram are dominant.
However, this type diagram turns to subdominant for the heavy vector mesons.

The direct contributions are very different from the indirect contributions, and they can only be calculated in a power series in
$(m_q/m_{h^0})^2$ or $(\Lambda_{QCD}/m_{h^0})^2$.  $m_q$ is the constituent quark mass in the meson, while $\Lambda_{QCD}$ represents the
hadronic scale. If the vector meson in the final state is longitudinally polarized, the direct contributions are produced from
subleading-twist projections leading to power suppressed. For the transversely polarized vector meson, leading-twist
projections provide direct contributions. With the asymptotic function $\phi_V^\perp(x)=6x(1-x)$\cite{GZBHS}, the direct contributions are obtained
\begin{eqnarray}
   F_{\perp\,{\rm direct}}^{VZ}
   &= \sum_q f_V^{q\perp} v_q\,\kappa_q\,\frac{3m_q}{2m_V}\,\frac{1-r_Z^2+2r_Z\ln r_Z}{(1-r_Z)^2} \,, \\
   \widetilde F_{\perp,{\rm direct}}^{VZ}
   &= \sum_q f_V^{q\perp} v_q\,\tilde\kappa_q\,\frac{3m_q}{2m_V}\,\frac{1-r_Z^2+2r_Z\ln r_Z}{(1-r_Z)^2} \,.
\end{eqnarray}
At first sight, this type direct contributions seem comparable with the indirect contributions in Eq.(\ref{FVZindirect}).
In fact, the numerical results of the direct contributions are still strongly suppressed.

\section{numerical results}

In this section, we calculate the numerical results and consider the constraints from the Higgs boson mass and Higgs boson decays
 $h^0\rightarrow \gamma\gamma$ and $h^0\rightarrow VV^*, V=(Z,W)$. The studied processes are $h^0\rightarrow Z \gamma$ and $h^0\rightarrow m_VZ$
with $m_V$ denoting $\rho, \omega, \phi, J/\psi$ and $\Upsilon$.
The used parameters in the BLMSSM are collected here
\begin{eqnarray}
&&m_{\tilde{Q}_3}=m_{\tilde{U}_3}=m_{\tilde{D}_3}=1.5{\rm TeV},~\lambda_d=0.4,~
\upsilon_{L_t}=3{\rm TeV},~A_{n_4}=A_{n_5}=1{\rm TeV},\nonumber\\&&
A_u=A_c=A_t=A_d=A_s=A_b=-1{\rm TeV},~~\tan\beta{_B}=0.4,~~B_4=L_4={3\over2},
\nonumber\\&&A_e=A_\mu=A_\tau=A'_e=A'_\mu=A'_\tau=500{\rm GeV},~~A_{BQ}=A_{BU}=A_{BD}=1{\rm TeV},
\nonumber\\&&
A'_u=A'_c=A'_t=A'_d=A'_s=A'_b=500{\rm GeV}, ~~~\tan\beta_{L}=3.9,~~~Y_{u_4}=0.8Y_t,
\nonumber\\
&&m_{\tilde{L}_4}=m_{\tilde{\nu}_4}=m_{\tilde{E}_4}=m_{\tilde{L}_5}=m_{\tilde{\nu}_5}
=m_{\tilde{E}_5}=1{\rm TeV},~~~m_{e_4}=m_{e_5}=100{\rm GeV},
\nonumber\\&&
m_{\nu_4}=m_{\nu_5}=90{\rm GeV},~~\mu_B=500{\rm GeV},~~g_B=1/3,~~\tan\beta=1.3,~~g_L=1/6,\nonumber\\&&m_{\tilde{Q}_1}
=m_{\tilde{U}_1}=m_{\tilde{D}_1}=m_{\tilde{Q}_2}=m_{\tilde{U}_2}=m_{\tilde{D}_2}=3{\rm TeV},~~~A_{e_4}=A_{e_5}=1.3{\rm TeV}.
\end{eqnarray}
To simplify the numerical discussion, we use the following relations
\begin{eqnarray}
&&A_{u_4}=A_{u_5}=A_{d_4}=A_{d_5}=AQ_{45},~~
~m_{\tilde{L}_1}=m_{\tilde{L}_2}=m_{\tilde{R}_1}=m_{\tilde{R}_2}=ML_S,\nonumber\\&&
m_{\tilde{U}_4}=m_{\tilde{D}_4}=m_{\tilde{Q}_4}=m_{\tilde{Q}_5}=m_{\tilde{U}_5}
=m_{\tilde{D}_5}=MQ_{45}.
\end{eqnarray}

\subsection{ the process $ h^0 \rightarrow m_V Z $}
At first, we study the decay $h^0\rightarrow \rho Z$ and the used parameters for the meson $\rho$ are
$m_\rho=0.77 {\rm GeV},~f_\rho=0.216 {\rm GeV},~Q_\rho=\frac{1}{\sqrt{2}},~v_\rho=\frac{1}{\sqrt{2}}(\frac{1}{2}-s_W^2)$ and
 $f^{\perp}_{\rho}/f_{\rho}=f^{q\perp}_{\rho}/f^q_{\rho}=0.72$. To obtain the numerical results, we suppose
$MQ_{45}=1{\rm TeV},~\lambda_u=0.4,~Y_{d_4}=Y_{d_5}=0.5Y_b,~AQ_{45}=1100{\rm GeV},
~ML_{S}=\upsilon_{B}=3{\rm TeV}$. $Y_b$ is the Yukawa coupling constant of the bottom quark
$Y_b = \sqrt{2} m_b/(\upsilon \cos\beta)$.

\begin{figure}[h]
\setlength{\unitlength}{1mm}
\centering
\includegraphics[width=2.9in]{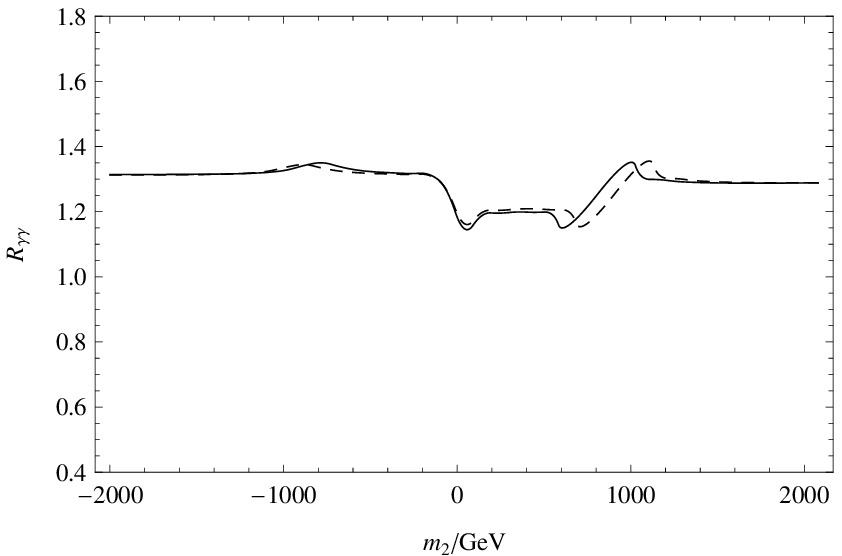},~~~~\includegraphics[width=2.9in]{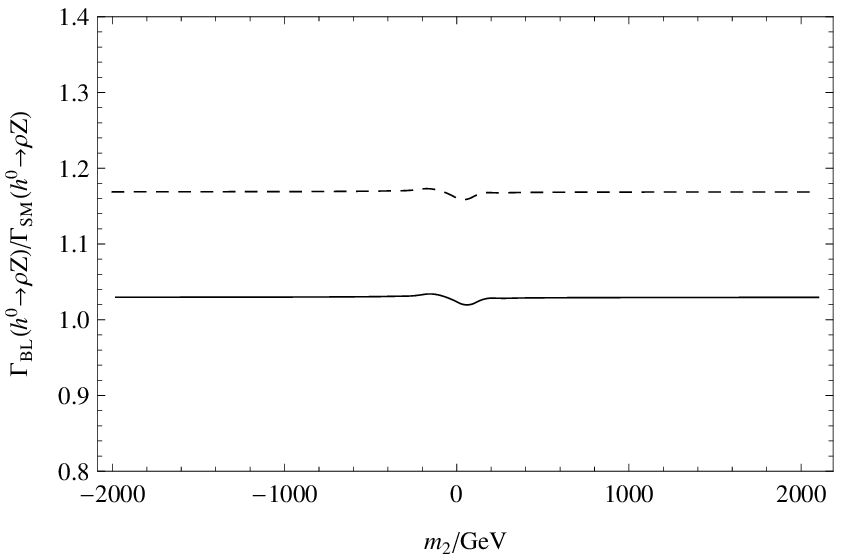}
\caption[]{With $Y_{u_5}=0.1Y_t$ and $\lambda_Q=0.4$, the results are plotted versus $m_2$.
The left diagram represents the $R_{\gamma\gamma}$ and the right diagram
shows the ratio $\Gamma_{BL}(h^0\rightarrow \rho Z)/\Gamma_{SM}(h^0\rightarrow \rho Z)$.
For the two diagrams, the solid and dashed lines correspond to $\mu=-800$ GeV and $\mu=-900$ GeV respectively.}\label{rhom2tu}
\end{figure}

$m_2$ can influence the results through the chargino contributions.  With $Y_{u_5}=0.1Y_t$ and $\lambda_Q=0.4$,
 $R_{\gamma\gamma}$ and the ratio $\Gamma_{BL}(h^0\rightarrow \rho Z)/\Gamma_{SM}(h^0\rightarrow \rho Z)$
versus $m_2$ are plotted respectively by the left and right diagrams in FIG.\ref{rhom2tu}. Here, $\Gamma_{BL}(h^0\rightarrow \rho Z)$ is the decay width
of the process $h^0\rightarrow \rho Z$ calculated in the BLMSSM. While, $\Gamma_{SM}(h^0\rightarrow \rho Z)$
represents the SM prediction of the decay $h^0\rightarrow \rho Z$.
$Y_t$ is the top quark Yukawa coupling constant $Y_t = \sqrt{2} m_t/(\upsilon \sin\beta)$. Using $\mu=-800$ GeV, we plot the results by the
solid line. While the dashed line corresponds to the results with  $\mu=-900$ GeV.
For $R_{\gamma\gamma}$, the solid line and the dashed line are almost overlapped.
In the $m_2$ region (-2000,~2000)GeV, $R_{\gamma\gamma}$ varies from 1.1 to 1.35.
The ratios $R_{VV}(V=Z,W)$ versus $m_2$ are very stable and near 1.15, which are not plotted here.
In our program, $m_{h^0}=125.1$ GeV is used as an input parameter. Therefore, the used parameters satisfy
the constraints from Higgs experiments. For the decay $h^0\rightarrow \rho Z$, the ratio $\Gamma_{BL}(h^0\rightarrow \rho Z)/\Gamma_{SM}(h^0\rightarrow \rho Z)$
varies weakly with $m_2$. In the right diagram of FIG.\ref{rhom2tu}, the solid line is around 1.03 and the dashed line is about 1.18.
In this process, the new physics contributions can almost reach $18\%$, which is considerable.

\begin{figure}[h]
\setlength{\unitlength}{1mm}
\centering
\includegraphics[width=3.3in]{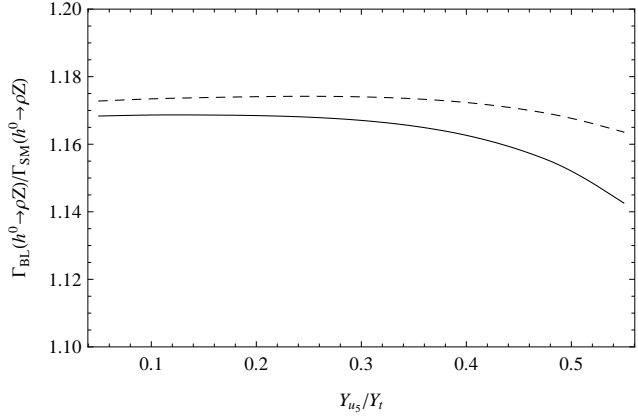}
\caption[]{With $\mu=-900$ GeV and $m_2=1500$ GeV,
the ratios $\Gamma_{BL}(h^0\rightarrow \rho Z)/\Gamma_{SM}(h^0\rightarrow \rho Z)$
versus $Y_{u_5}$ are plotted by the dashed line ($\lambda_{Q}=0.8$) and solid line ($\lambda_{Q}=0.4$) respectively.
}\label{rhoYu5tu}
\end{figure}

Because the used parameters satisfy the Higgs experiment constraints, we do not show the results for $R_{\gamma\gamma}$ and
$R_{VV} (V=Z,W)$ again in the following numerical discussions. $Y_{u_5}$ has relation with exotic quark and exotic squark,
so  $Y_{u_5}$  influences the new physics corrections. In FIG.\ref{rhoYu5tu},
 with  $\mu=-900$ GeV, $m_2=1500$ GeV, the ratios versus $Y_{u_5}$ are plotted by
the dashed line$(\lambda_{Q}=0.8)$ and solid line$(\lambda_{Q}=0.4)$ respectively. The solid line
and dashed line are both decreasing functions, when $Y_{u_5}$ varies from $0.1Y_t$ to $0.5Y_t$.
 The dashed line is around 1.17
and up the solid line. At the same time, the biggest value of the solid line is around 1.168 with $Y_{u_5}$ near $0.1Y_t$,
and its smallest value is about 1.145 with $Y_{u_5}$ near $0.5Y_t$.
Generally speaking, corresponding to the SM results the new physics contributions are approximately $17\%$ in this condition.

\begin{figure}[h]
\setlength{\unitlength}{1mm}
\centering
\includegraphics[width=3.3in]{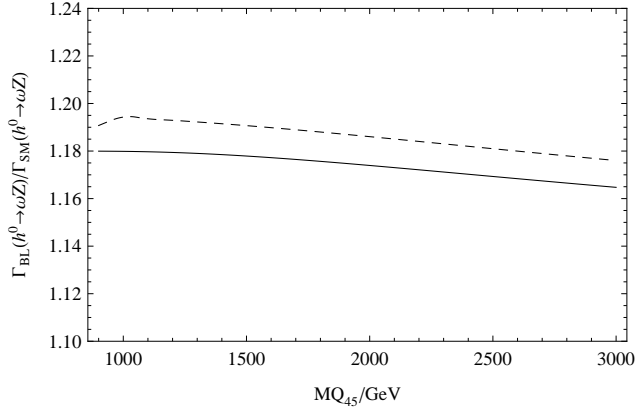}
\caption[]{With $\lambda_u=0.4$ and $Y_{d_4}=Y_{d_5}=0.5Y_b$,
 the ratio $\Gamma_{BL}(h^0\rightarrow \omega Z)/\Gamma_{SM}(h^0\rightarrow \omega Z)$
versus $MQ_{45}$ is plotted by the dashed line($\upsilon_{B}=4$ TeV) and solid line($\upsilon_{B}=3$ TeV) respectively.
}\label{omegamQ45tu}
\end{figure}

Secondly, the process $h^0\rightarrow \omega Z$ is calculated with the  meson $\omega$ parameters,
$m_\omega=0.782{\rm GeV},~f_\omega=0.194{\rm GeV},~Q_\omega=\frac{1}{3\sqrt{2}},~v_\omega=-\frac{s_W^2}{3\sqrt{2}}$ and $f^{\perp}_{\omega}/f_{\omega}=f^{q\perp}_{\omega}/f^q_{\omega}=0.71$\cite{decayconstant}. Here, some suppositions are taken as
$AQ_{45}=1100{\rm GeV},~\mu=-900{\rm GeV},~m_2=1500{\rm GeV},~Y_{u_5}=0.1Y_t,~\lambda_Q=0.4,~
ML_{S}=3{\rm TeV}$. The mass squared matrices of the exotic squarks have the elements $MQ_{45}$.
Therefore, the ratio $\Gamma_{BL}(h^0\rightarrow \omega Z)/\Gamma_{SM}(h^0\rightarrow \omega Z)$ versus $MQ_{45}$
is researched numerically as $\lambda_u=0.4$ and $Y_{d_4}=Y_{d_5}=0.5Y_b$.
In FIG.\ref{omegamQ45tu}, the dashed line is obtained with $\upsilon_{B}=4$ TeV, and the solid line
represents the results gotten with $\upsilon_{B}=3$ TeV.
$\upsilon_{B}$
affects the masses of exotic quark and exotic squark, and influences the ratio.
From FIG.\ref{omegamQ45tu}, it is obviously that both the solid line and the dashed line turn small weakly with the increasing
$MQ_{45}$ as $MQ_{45}>1000$ GeV. This character is easy to understand, because large $MQ_{45}$ leads to heavy exotic quark (squark)
and suppresses their contributions. Both the solid line and the dashed line are in the region $1.17\sim1.20$.

\begin{figure}[h]
\setlength{\unitlength}{1mm}
\centering
\includegraphics[width=3.3in]{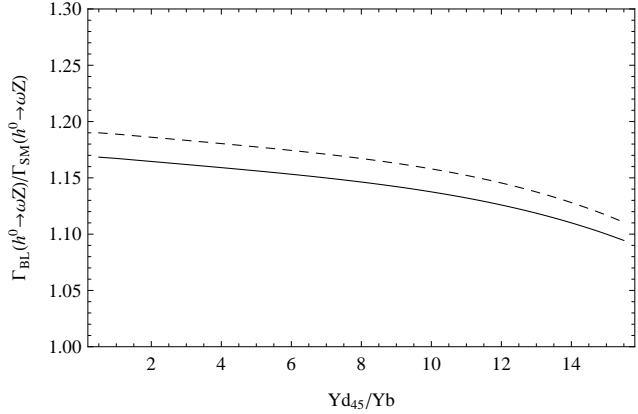}
\caption[]{With $\upsilon_{B}=3$ TeV and $MQ_{45}=1$ TeV, the ratio $\Gamma_{BL}(h^0\rightarrow \omega Z)/\Gamma_{SM}(h^0\rightarrow \omega Z)$
versus $Yd_{45}$ is plotted by the dashed line$(\lambda_{u}=0.5)$ and solid line$(\lambda_{u}=0.3)$ respectively.}
\label{omegaYd45tu}
\end{figure}

 The Yukawa couplings $Y_{d_4}$ and $Y_{d_5}$ for the exotic quarks are important parameters.
 With $\upsilon_{B}=3$ TeV and $MQ_{45}=1$ TeV, the effects of $Y_{d_4}=Y_{d_5}=Yd_{45}$ to
 the ratio $\Gamma_{BL}(h^0\rightarrow \omega Z)/\Gamma_{SM}(h^0\rightarrow \omega Z)$ are researched in FIG.\ref{omegaYd45tu}
 and shown by the solid line($\lambda_u=0.5$) and dashed line($\lambda_u=0.3$). Both the solid line and dashed line
 turn small in the $Yd_{45}$ region ($0.5Y_b\sim 15 Y_b$). The biggest value and the smallest value of the dashed line are
 1.19 and 1.13 respectively. The values of the solid line vary from 1.17 to 1.10. With the same value of $Yd_{45}$, the dashed line
 is about 0.02 bigger than the solid line.

\begin{figure}[h]
\setlength{\unitlength}{1mm}
\centering
\includegraphics[width=3.3in]{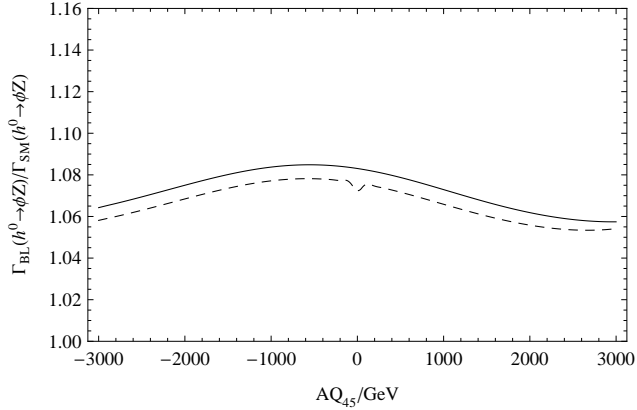}
\caption[]{With $\lambda_Q=0.4$ and $\lambda_u=0.5$, the ratio $\Gamma_{BL}(h^0\rightarrow \phi Z)/\Gamma_{SM}(h^0\rightarrow \phi Z)$
versus $AQ_{45}$ is plotted by the dashed line. While the solid line corresponds to the ratio for $\lambda_Q=0.8$ and $\lambda_u=0.3$.}
\label{phiAud45tu}
\end{figure}

Thirdly, the needed constants for $\phi$ are
 $m_\phi=1.02{\rm GeV},~f_\phi=0.223{\rm GeV},~Q_\phi=-\frac{1}{3},~v_\phi=-\frac{1}{4}+\frac{s_W^2}{3}$
 and $f^{\perp}_{\phi}/f_{\phi}=f^{q\perp}_{\phi}/f^q_{\phi}=0.76$. Supposing
 $Y_{u_5}=0.1Y_t,~\lambda_Q=0.4,~ MQ_{45}=1{\rm TeV}, ~\upsilon_{B}=ML_{S}=3{\rm TeV},~
m_2=1500{\rm GeV},\mu=-900 {\rm GeV}$ and $Y_{d_4}=Y_{d_5}=0.5Y_b$, we research the decay $h^0\rightarrow \phi Z$ numerically.
The gotten results for the ratio $\Gamma_{BL}(h^0\rightarrow \phi Z)/\Gamma_{SM}(h^0\rightarrow \phi Z)$
versus $AQ_{45}$ are plotted in FIG.\ref{phiAud45tu},
in which the dashed line is obtained with $\lambda_Q=0.4$ and $\lambda_u=0.5$.
While, the solid line represents the results as $\lambda_Q=0.8$ and $\lambda_u=0.3$. The solid line and the dashed line
are of the same behavior versus $AQ_{45}$, and they are very near. The highest point of the solid line is about 1.08 and gotten
 around the point $AQ_{45}=-500$ GeV.
The $AQ_{45}$ effects to the ratios are small, and the results are in the range $1.05\sim1.08$.

The quark constituent of $J/\psi$ is $c\bar{c}$, whose parameters used here are
$m_{J/\psi}=3.097{\rm GeV},~
f_{J/\psi}=0.403{\rm GeV},~Q_{J/\psi}=\frac{2}{3},~v_{J/\psi}=\frac{1}{4}-\frac{2s_W^2}{3}$ and
 $f^{\perp}_{J/\psi}/f_{J/\psi}=f^{q\perp}_{J/\psi}/f^q_{J/\psi}=0.91$.
To get the numerical results of the decay $h^0\rightarrow J/\psi Z$, the parameters are taken as
$m_2=1500{\rm GeV},~\mu=-900{\rm GeV},~Y_{u_5}=0.1Y_t,~Y_{d_5}=0.5Y_b,~\lambda_Q=0.4,~AQ_{45}=1100{\rm GeV},~\lambda_u=0.4,~MQ_{45}=1{\rm TeV}$ . We study the ratio
$\Gamma_{BL}(h^0\rightarrow J/\psi Z)/\Gamma_{SM}(h^0\rightarrow J/\psi Z)$ versus  $ML_{S}$ and $\upsilon_{B}$. The obtained numerical results
are stable and around 1.08. That is to say the effects from $ML_{S}$ and $\upsilon_{B}$ are tiny to the decay $h^0\rightarrow J/\psi Z$.

At last, $h^0\rightarrow \Upsilon Z$ is studied with the following parameters from the meson: $m_{\Upsilon}=9.46{\rm GeV},~f_{\Upsilon}=0.684{\rm GeV},
~Q_{\Upsilon}=-\frac{1}{3},~v_{\Upsilon}=-\frac{1}{4}+\frac{s_W^2}{3},
~f^{\perp}_{\Upsilon}/f_{\Upsilon}=f^{q\perp}_{\Upsilon}/f^q_{\Upsilon}=1.09$.
 As the heavy vector meson, $\Upsilon$ is made up of $b\bar{b}$. In the numerical calculation, we try to adjust many
 BLMSSM parameters, but the obtained  ratios for $\Gamma_{BL}(h^0\rightarrow \Upsilon Z)/\Gamma_{SM}(h^0\rightarrow \Upsilon Z)$
 are all very near 1. For the decay $h^0\rightarrow \Upsilon Z$, the new physics corrections
 are very small and negligible.

\subsection{the process $h^0\rightarrow Z\gamma$}
  The new physics contributions to the decay $h^0\rightarrow m_V Z $ come from the effective coupling of $h^0Z\gamma$.
  Furthermore, considering the LHC results of $h^0\rightarrow \gamma\gamma$ and $h^0\rightarrow ZZ^*$,
  researching the process $h^0\rightarrow Z \gamma$ turns more important to identify the nature of $h^0$.
  The values of the ratios $R_{\gamma\gamma}$ and $R_{ZZ}$ are
$R_{\gamma\gamma}=1.16\pm0.18$ and $R_{ZZ}=1.29^{+0.26}_{-0.23}$ respectively. Therefore, the value of $R_{Z\gamma}$ is of great interest.

\begin{figure}[h]
\setlength{\unitlength}{1mm}
\centering
\includegraphics[width=3.3in]{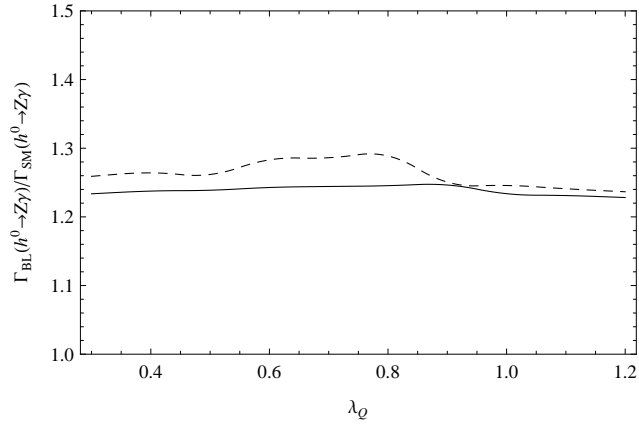}
\caption[]{With $\lambda_u=0.5$,
we plot the ratio $\Gamma_{BL}(h^0\rightarrow Z\gamma)/\Gamma_{SM}(h^0\rightarrow Z\gamma)$
versus $\lambda_Q$ by the solid line($\upsilon_B=3$ TeV) and dashed line($\upsilon_B=4.3$ TeV) respectively.}
\label{hzglamQ}
\end{figure}

\begin{figure}[h]
\setlength{\unitlength}{1mm}
\centering
\includegraphics[width=3.3in]{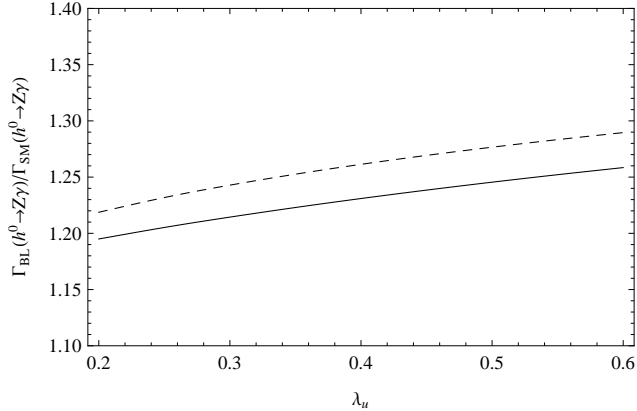}
\caption[]{With $\lambda_Q=0.8$,
we plot the ratio $\Gamma_{BL}(h^0\rightarrow Z\gamma)/\Gamma_{SM}(h^0\rightarrow Z\gamma)$
versus $\lambda_u$ by the solid line($\upsilon_B=3$ TeV) and dashed line($\upsilon_B=4$ TeV) respectively.}
\label{hzglamu}
\end{figure}

 In the numerical calculation of the process $h^0\rightarrow Z\gamma$, we use the parameters as
$m_2=1500{\rm GeV},~\mu=-900{\rm GeV},~Y_{u_5}=0.1Y_t,~Y_{d_5}=0.5Y_b,~AQ_{45}=1100{\rm GeV},~MQ_{45}=1{\rm TeV}.$
$\lambda_Q$ and $\lambda_u$ are coupling constants of exotic quarks and $\Phi_B(\varphi_B)$, which are both important parameters.
The ratio $\Gamma_{BL}(h^0\rightarrow Z\gamma)/\Gamma_{SM}(h^0\rightarrow Z\gamma)$
versus $\lambda_Q$ is plotted in FIG.\ref{hzglamQ} as $\lambda_u=0.5$.
The solid line and dashed line correspond to the results with $\upsilon_B=3$ TeV and $\upsilon_B=4.3$ TeV. The solid line is around
1.24 and the dashed line can almost reach 1.30 when $\lambda_Q$ is near 0.8.
The results varying with $\lambda_u$ are also calculated. Taking $\lambda_Q=0.8$, we
 plot the ratio $\Gamma_{BL}(h^0\rightarrow Z\gamma)/\Gamma_{SM}(h^0\rightarrow Z\gamma)$
versus $\lambda_u$ by the solid line ($\upsilon_{B}=3$ TeV) and dashed line ($\upsilon_{B}=4$ TeV) in FIG.\ref{hzglamu}.
The dashed line is up the solid line and they are both increasing functions in the $\lambda_u$ region $(0.2\sim0.6)$.
The values of the solid line are from 1.2 to 1.26. At the same time, the dashed line varies from 1.22 to 1.29.
In the whole, comparing with the SM results, the new physics corrections to $h^0\rightarrow Z\gamma$ are $23\sim30\%$.

\section{discussion and conclusion}
In this work, we study the decays $h^0\rightarrow Z\gamma$ and $h^0\rightarrow m_VZ$ with $m_V=\rho,\omega,\phi,J/\psi,\Upsilon$ in the BLMSSM.
The decay $h^0\rightarrow m_VZ$ has two type contributions: the direct contributions and the indirect contributions.
In the direct contributions, the quarks coupling with Higgs make up of the
final state vector meson directly. For the indirect contributions, the Higgs couples to the on-shell $Z$ and an off-shell gauge boson($\gamma$ or $Z$).
The off-shell boson  converts into the vector meson in the end.
As discussed in the Ref.\cite{htomz}, the indirect contributions are more important than the direct contributions.
In the SM, there is $h^0ZZ$ coupling at tree level and the $h^0\gamma Z$ coupling is produced from loop diagrams. In the models beyond SM, there
can be CP-even coupling constant $C_{\gamma Z}$ and CP-odd coupling constant $\tilde{C}_{\gamma Z}$.
Even for the new physics, the CP-even part is more important than the CP-odd part. We use the
effective Lagragian method to calculate the effective constants $C_{\gamma Z}$ and $\tilde{C}_{\gamma Z}$ for the vertex $h^0\gamma Z$.

The experiment results of the ratios $R_{\gamma\gamma}$ and $R_{ZZ}$,  are
$R_{\gamma\gamma}=1.16\pm0.18$ and $R_{ZZ}=1.29^{+0.26}_{-0.23}$.
 Our numerical results of the ratio $\Gamma_{BL}(h^0\rightarrow Z\gamma)/\Gamma_{SM}(h^0\rightarrow Z\gamma)$ are around 1.29.
For $h^0\rightarrow ZZ^*,~h^0\rightarrow\gamma\gamma$ and $h^0\rightarrow Z\gamma$,
the one loop diagrams are similar. Different from $h^0\rightarrow\gamma\gamma$ and $h^0\rightarrow Z\gamma$, the process $h^0\rightarrow ZZ^*$ has tree level coupling
of $h^0ZZ$, and the neutral particles can appear in the one loop diagrams.
 $h^0\rightarrow\gamma\gamma$ and $h^0\rightarrow Z\gamma$
 are only produced from loop diagrams
that are almost same, with the differences just coming from
the couplings of the exchanged charged particles with $Z$ and $\gamma$.
Generally speaking, these three processes are very similar. Taking into account the values of the ratios
$R_{\gamma\gamma}$ and $R_{ZZ}$,  the ratio $\Gamma_{BL}(h^0\rightarrow Z\gamma)/\Gamma_{SM}(h^0\rightarrow Z\gamma)$ near 1.3 is reasonable, and
should not be very large. The process $h^0\rightarrow Z\gamma$ is comparable with $h^0\rightarrow \gamma\gamma(ZZ^*)$ and should match experimental sensitivities.

From the numerical analysis, one finds that the new physics contributions to the decays
$h^0\rightarrow \rho Z$ and $h^0\rightarrow \omega Z$ are about $15\%\sim20\%$. While the new physics corrections to the
processes $h^0\rightarrow \phi Z$ and $h^0\rightarrow J/\psi Z$ are during $5\%\sim10\%$. $\Upsilon$ is the heaviest one in our studied vector mesons$(\rho,~\omega,~\phi,~J/\psi,~\Upsilon)$.
We try to adjust many parameters to make the new physics corrections to $h^0\rightarrow \Upsilon Z$ considerable.
However, we have to
admit that comparing with the SM results, the corrections from the new physics are tiny.
Our results imply an interesting law: for the decay $h^0\rightarrow m_V Z$,
 new physics contributions of the BLMSSM are large when the final state meson is light.
This law is consistent with that the indirect contributions from the effective coupling $h^0\gamma Z$
are important for light vector mesons as discussed by the authors\cite{htomz}.
Considering $\Gamma_{BL}(h^0\rightarrow Z\gamma)/\Gamma_{SM}(h^0\rightarrow Z\gamma)\sim1.30$,
it is not small that the new physics corrections to the rare decays $h^0\rightarrow m_V Z$ reach $15\sim20\%$
of the SM predictions. This work is valuable for the experimentalists to
detect the decays $h^0\rightarrow Z\gamma$ and $h^0\rightarrow m_V Z $.

In the models beyond SM, the new physics can produce P-odd coupling and give corrections to P-even coupling.
Therefore, the rate of $h^0\rightarrow m_VZ$ is enhanced. $h^0\rightarrow m_VZ$ is rare decay, whose branching ratio is small.
 So the present experiments are unable to measure $h^0\rightarrow m_VZ$. The decay $h^0\rightarrow m_VZ$ is theoretically calculable,
 experimentally promising, and should be a priority at future hadron colliders. It is accessible at HL-LHC and
 future high energy colliders\cite{HQQ2}.

{\bf Acknowledgments}

This work is supported by the National Natural Science Foundation
of China (NNSFC) under contract
No. 11535002, No. 11605037, No. 11705045,
the Natural Science Foundation of Hebei province with Grant
No. A2016201010 and No. A2016201069, Hebei Key Lab of Optic-Electronic Information and
Materials, the midwest universities comprehensive strength
promotion project. At last, thanks professor Xue-Qian Li very much for English revision.
\appendix
\section{}
The couplings between the lightest neutral CP-even Higgs $h^0$ and the exotic sleptons are collected here.
 \begin{eqnarray}
  &&\mathcal{L}_{h^{0}\tilde{L}^\prime \tilde{L}^\prime}=\sum_{i,j=1}^2\tilde{E}^{\prime i*}_4\tilde{E}^{\prime j}_4
  h^0[(E^u_4)_{ij}\cos\alpha-(E^d_4)_{ij}\sin\alpha]\nonumber\\&&\hspace{1.4cm}+\sum_{i,j=1}^2\tilde{E}^{\prime i*}_5\tilde{E}^{\prime j}_5
  h^0[(E^u_5)_{ij}\cos\alpha-(E^d_5)_{ij}\sin\alpha].\label{LHLpLp}
  \end{eqnarray}
  The concrete forms of the couplings $(E^u_4)_{ij}, (E^d_4)_{ij},(E^u_5)_{ij},(E^d_5)_{ij}$ read as
  \begin{eqnarray}
  &&  (E^u_4)_{ij}=-e^2\upsilon\sin\beta(\frac{1}{2c_W^2}\delta_{ij}+\frac{1-4s_W^2}{4s_W^2c_W^2}(Z_{\tilde{e}_4}^\dag)^{i1}Z_{\tilde{e}_4}^{1j})
  -\frac{\mu^*}{\sqrt{2}}Y_{e_4}(Z_{\tilde{e}_4}^\dag)^{i2}Z_{\tilde{e}_4}^{1j},\nonumber\\&&
  (E^d_4)_{ij}=e^2\upsilon\cos\beta(\frac{1}{2c_W^2}\delta_{ij}+\frac{1-4s_W^2}{4s_W^2c_W^2}(Z_{\tilde{e}_4}^\dag)^{i1}Z_{\tilde{e}_4}^{1j})
  -\upsilon\cos\beta|Y_{e_4}|^2\delta_{ij}-\frac{A_{E_4}}{\sqrt{2}}(Z_{\tilde{e}_4}^\dag)^{i2}Z_{\tilde{e}_4}^{1j},\nonumber\\&&
  (E^u_5)_{ij}=e^2\upsilon\sin\beta(\frac{1}{2c_W^2}\delta_{ij}+\frac{1-4s_W^2}{4s_W^2c_W^2}(Z_{\tilde{e}_5}^\dag)^{i2}Z_{\tilde{e}_5}^{2j})
  -\upsilon\sin\beta|Y_{e_5}|^2\delta_{ij}-\frac{A_{E_5}}{\sqrt{2}}(Z_{\tilde{e}_5}^\dag)^{i2}Z_{\tilde{e}_5}^{1j},\nonumber\\&&
  (E^d_5)_{ij}=-e^2\upsilon\cos\beta(\frac{1}{2c_W^2}\delta_{ij}+\frac{1-4s_W^2}{4s_W^2c_W^2}(Z_{\tilde{e}_4}^\dag)^{i2}Z_{\tilde{e}_4}^{2j})
  -\frac{\mu^*}{\sqrt{2}}Y_{e_5}(Z_{\tilde{e}_5}^\dag)^{i2}Z_{\tilde{e}_5}^{1j}.\label{LHLpLp2}
  \end{eqnarray}

In the mass basis, the couplings between $h^0$ and exotic squarks are
\begin{eqnarray}
&&{\cal L}_{h^0\tilde{\cal U}_i^*\tilde{\cal U}_j}=\sum\limits_{i,j}^4\Big\{\Big[\xi_{uij}^S\cos\alpha
-\xi_{dij}^S\sin\alpha\Big]h^0\tilde{\cal U}_i^*\tilde{\cal U}_j+\Big[\eta_{uij}^S\cos\alpha
-\eta_{dij}^S\sin\alpha\Big]h^0\tilde{\cal D}_i^*\tilde{\cal D}_j,
\end{eqnarray}
with
\begin{eqnarray}
&&\xi_{uij}^S={1\over\sqrt{2}}Y_{u_5}\mu\Big(U_{i3}^\dagger U_{4j}+U_{i4}^\dagger U_{3j}\Big)
+{1\over2}\lambda_{Q}Y_{u_4}\upsilon_{B}\Big(U_{i3}^\dagger U_{2j}+U_{i2}^\dagger U_{3j}\Big)
\nonumber\\
&&\hspace{1.2cm}
-{1\over2}\lambda_{u}Y_{u_4}\overline{\upsilon}_{B}\Big(U_{i1}^\dagger U_{4j}+U_{i4}^\dagger U_{1j}\Big)
+{e^2\over4s_{\rm W}^2}\upsilon_{u}\Big(U_{i3}^\dagger U_{3j}-U_{i1}^\dagger U_{1j}\Big)
\nonumber\\
&&\hspace{1.2cm}
+{e^2\upsilon_{u}\over12c_{\rm W}^2}\Big(U_{i1}^\dagger U_{1j}-U_{i3}^\dagger U_{3j}
-4U_{i2}^\dagger U_{2j}+4U_{i4}^\dagger U_{4j}\Big)
-{A_{u_4}Y_{u_4}\over\sqrt{2}}\Big(U_{i2}^\dagger U_{1j}+U_{i1}^\dagger U_{2j}\Big)
\;,\nonumber\\
&&\xi_{dij}^S={1\over\sqrt{2}}Y_{u_4}\mu\Big(U_{i2}^\dagger U_{1j}+U_{i1}^\dagger U_{2j}\Big)
+{1\over2}\lambda_{Q}Y_{u_5}\upsilon_{B}\Big(U_{i5}^\dagger U_{1j}+U_{i1}^\dagger U_{5j}\Big)
\nonumber\\
&&\hspace{1.2cm}
-{1\over2}\lambda_{u}Y_{u_5}\overline{\upsilon}_{B}\Big(U_{i2}^\dagger U_{3j}+U_{i3}^\dagger U_{2j}\Big)
-{e^2\over4s_{\rm W}^2}\upsilon_{d}\Big(U_{i3}^\dagger U_{3j}+U_{i1}^\dagger U_{1j}\Big)
\nonumber\\
&&\hspace{1.2cm}
-{e^2\upsilon_{d}\over12c_{\rm W}^2}\Big(U_{i1}^\dagger U_{1j}-U_{i3}^\dagger U_{3j}
-4U_{i2}^\dagger U_{2j}+4U_{i4}^\dagger U_{4j}\Big)
-{A_{u_5}Y_{u_5}\over\sqrt{2}}\Big(U_{i3}^\dagger U_{4j}+U_{i4}^\dagger U_{3j}\Big)
\;,\nonumber\\
&&\eta_{uij}^S={1\over\sqrt{2}}Y_{d_4}\mu\Big(D_{i2}^\dagger D_{1j}+D_{i1}^\dagger D_{2j}\Big)
+{1\over2}\lambda_{Q}Y_{d_5}\upsilon_{B}\Big(D_{i4}^\dagger D_{1j}+D_{i1}^\dagger D_{4j}\Big)
\nonumber\\
&&\hspace{1.2cm}
-{1\over2}\lambda_{d}Y_{d_5}\overline{\upsilon}_{B}\Big(D_{i2}^\dagger D_{3j}+D_{i3}^\dagger D_{2j}\Big)
+{e^2\over4s_{\rm W}^2}\upsilon_{u}\Big(D_{i1}^\dagger D_{1j}-D_{i3}^\dagger D_{3j}\Big)
\nonumber\\
&&\hspace{1.2cm}
+{e^2\upsilon_{u}\over12c_{\rm W}^2}\Big(D_{i1}^\dagger D_{1j}-D_{i3}^\dagger D_{3j}
+2D_{i2}^\dagger D_{2j}-2D_{i4}^\dagger D_{4j}\Big)
-{A_{d_5}Y_{d_5}\over\sqrt{2}}\Big(D_{i3}^\dagger D_{4j}+D_{i4}^\dagger D_{3j}\Big)
\;,\nonumber\\
&&\eta_{dij}^S={1\over\sqrt{2}}Y_{d_5}\mu\Big(D_{i3}^\dagger D_{4j}+D_{i4}^\dagger D_{3j}\Big)
+{1\over2}\lambda_{Q}Y_{d_4}\upsilon_{B}\Big(D_{i3}^\dagger D_{2j}+D_{i2}^\dagger D_{3j}\Big)
\nonumber\\
&&\hspace{1.2cm}-{e^2\upsilon_{u}\over12c_{\rm W}^2}\Big(D_{i1}^\dagger D_{1j}-D_{i3}^\dagger D_{3j}
+2D_{i2}^\dagger D_{2j}-2D_{i4}^\dagger D_{4j}\Big)
-{e^2\over4s_{\rm W}^2}\upsilon_{d}\Big(D_{i1}^\dagger D_{1j}-D_{i3}^\dagger D_{3j}\Big)
\nonumber\\
&&\hspace{1.2cm}
-{1\over2}\lambda_{d}Y_{d_4}\overline{\upsilon}_{B}\Big(D_{i1}^\dagger D_{4j}+D_{i4}^\dagger D_{1j}\Big)
-{A_{d_4}Y_{d_4}\over\sqrt{2}}\Big(D_{i2}^\dagger D_{1j}+D_{i1}^\dagger D_{2j}\Big)
.
\end{eqnarray}


\begin{thebibliography}{99}
\bibitem{higgs125}
CMS collaboration, Phys. Lett. B 716 (2012) 30; ATLAS collaboration,  Phys. Lett. B 716 (2012) 1.
\bibitem{neutrinomass}
T2K collaboration, Phys. Rev. Lett. 107 (2011) 041801; DAYA-BAY collaboration, Phys. Rev. Lett. 108 (2012) 171803.
\bibitem{2016pdg}Particle Data Group collaboration, Chin. Phys. C 40 (2016) 100001.

\bibitem{twoHiggs}
G.C. Branco, P.M. Ferreira, L. Lavoura et al., Phys. Rept. 516 (2012) 1;
D. Atwood, L. Reina, A. Soni, Phys. Rev. D55 (1997) 3156.
\bibitem{MSSM}
 J. Rosiek, Phys. Rev. D 41 (1990) 3464, hep-ph/9511250.
\bibitem{SUSY}
 H.P. Nilles, Phys. Rept. 110 (1984) 1; H.E. Haber and G.L. Kane, Phys. Rept. 117 (1985) 75.

\bibitem{vertex}D.N. Gao, Phys. Lett. B 737 (2014) 366; T. Modak and R. Srivastava, Mod. Phys. Lett. A 32 (2017) 1750004.
\bibitem{htozgexp} ATLAS Collaboration, Phys. Lett. B 732 (2014) 8-27; ATLAS Collaboration, JHEP 10 (2017) 112;
The CMS collaboration, JHEP 01 (2017) 076; The CMS Collaboration, Phys. Lett. B 772 (2017) 363-387.


\bibitem{htomgamma}
 A.L. Kagan, G. Perez, F. Petriello et al., Phys. Rev. Lett. 114 (2015) 101802;
G.T. Bodwin, H.S. Chung, J.H. Ee et al., Phys. Rev. D 90 (2014) 113010; M. Konig and M. Neubert, JHEP 08 (2015) 012.

\bibitem{HQQ1} B. Bhattacharya, A. Datta, D. London, Phys. Lett. B, 736 (2014) 421-427

\bibitem{HQQ2}A.L. Kagan, G. Perez, F. Petriello, Y. Soreq, S. Stoynev, J. Zupan, Phys. Rev. Lett. 114, 101802 (2015);
G. Isidori, A.V. Manohar, M. Trott, Phys. Lett. B 728 (2014) 131-135


\bibitem{htomz}
S. Alte, M. Koniga and M. Neubert, JHEP 12 (2016) 037.
\bibitem{indirect}G. Isidori, A.V. Manohar and M. Trott, Phys. Lett. B 728 (2014) 131;
M.G. Alonso and G. Isidori, Phys. Lett. B 733 (2014) 359.
\bibitem{other1}G.P. Lepage and S.J. Brodsky, Phys. Lett. B 87 (1979) 359; G.P. Lepage and S.J. Brodsky, Phys. Rev. D 22 (1980) 2157.
\bibitem{other2} A.V. Efremov and A.V. Radyushkin, Phys. Lett. B 94 (1980) 245; V.L. Chernyak and A.R. Zhitnitsky, Phys. Rept. 112 (1984) 173.
\bibitem{BLfirst} P.F. Perez, Phys. Lett. B 711 (2012) 353; J.M. Arnold, P.F. Perez, B. Fornal, S. Spinner, Phys. Rev. D 85
(2012) 115024; P.F. Perez, M.B. Wise, Phys. Rev. D 84 (2011) 055015.
\bibitem{TFBL}
T.F. Feng, S.M. Zhao, H.B. Zhang et al., Nucl. Phys. B 871 (2013) 223.
\bibitem{zhaoBL}S.M. Zhao, T.F. Feng, H.B. Zhang et al., Phys. Rev. D 92 (2015) 115016;
 S.M. Zhao, T.F. Feng, X.J. Zhan et al., JHEP 07 (2015) 124; S.M. Zhao, T.F. Feng, B. Yan et al., JHEP 10 (2013) 020.
\bibitem{CGZ}
L. Bergstrom and G. Hulth, Nucl. Phys. B 259 (1985) 137 [Erratum ibid. B 276 (1986) 744] .
\bibitem{PDG2014}Particle Data Group collaboration, Chin.
Phys. C 38 (2014) 090001.
\bibitem{SM14JHEP}S.M. Zhao, T.F. Feng, H.B. Zhang et al., JHEP 11 (2014) 119.
\bibitem{QCDCR}M. Spira, A. Djouadi and P.M. Zerwas, Phys.
Lett. B 276 (1992) 350.
\bibitem{UPbound} CMS collaboration, Phys. Lett. B 726 (2013) 587;
 ATLAS collaboration,  Phys. Lett. B 732 (2014) 8.
 \bibitem{GZBHS}V.L. Chernyak, A.R. Zhitnitsky, Nucl. Phys. B 201 (1996) 2182; N.H. Fuchs, M.D. Scadron, Phys. Rev. D 20 (1979) 2421;
 M. Beneke, G. Buchalla, M. Neubert, C.T. Sachrajda, Nucl. Phys. B 591 (2000) 313.
 \bibitem{decayconstant}Y. Grossman, M. Konig and M. Neubert, JHEP 04 (2015) 101.


\end{thebibliography}
\end{document}